\documentclass{elsart}

\usepackage[dvips]{graphicx}

\usepackage{amssymb}

\begin{document}
\begin{frontmatter}


\title{Precision Measurement of Orthopositronium Decay Rate Using 
\boldmath{${\rm SiO_2}$} Powder}
\author{O. Jinnouchi},
\author{S. Asai}, and
\author{T. Kobayashi}
\address{International Center for Elementary Particle Physics,\\
University of Tokyo, Faculty of Science Building 1, 7-3-1 Hongo,\\
Bunkyo-ku, Tokyo, 113-0033, Japan}


\begin{abstract}

The intrinsic decay rate of orthopositronium formed in ${\rm SiO_2}$ 
powder is measured using the direct 
$2\gamma$ correction method such that the time dependence of the 
pick-off annihilation rate is precisely 
determined using high energy-resolution germanium detectors. 
As a systematic test, two different types of 
${\rm SiO_2}$ powder are used with consistent findings. The intrinsic 
decay rate of orthopositronium is 
found to be $7.0396\pm0.0012 (stat.)\pm0.0011 (sys.)\mu s^{-1}$, 
which is consistent with previous 
measurements using ${\rm SiO_2}$ powder with about twice the accuracy. 
Results agree well with a 
recent $O(\alpha^2)$ QED prediction, varying $3.8-5.6$ experimental 
standard deviations from other 
measurements.  

\end{abstract}
\begin{keyword}
Positronium\sep bound states \sep QED
\PACS 36.10.Dr\sep 11.10.St
\end{keyword}
\end{frontmatter}
\section{Introduction}\label{sec:intro}

Positronium (Ps), the bound state of an electron and positron, is a 
pure quantum electro-dynamical system 
providing a highly sensitive field for testing bound state quantum 
electrodynamics (QED). The triplet 
(${\rm 1^3S_1}$) state of Ps, orthopositronium (o-Ps), will in most 
cases decay into three photons due to 
odd-parity under the C-transformation. Because o-Ps has a lifetime 
about 1140 times longer compared to 
the singlet state, parapositronium (p-Ps), this enables direct and 
precise measurement of the intrinsic decay 
rate of orthopositronium, $\lambda_{{\rm o}\mbox{-}{\rm Ps}}$ 
\cite{GAS89,CAV90}, although 
obtained values are much larger, i.e., 5.2 and 8.2 experimental 
standard deviations, than a recent non-relativistic QED calculation 
($7.039~979(11)~\mu s^{-1}$) corrected up to $O(\alpha^2)$ \cite{ADKINS-4}. 
To elucidate discrepancies, a variety of experiments have since 
been carried out to search for the exotic 
decay mode of o-Ps, resulting in no evidence so far 
\cite{EXOTIC-LL1,EXOTIC-LL0,EXOTIC-SL0,EXOTIC-SL1,EXOTIC-SL2,EXOTIC-UB0,EXOTIC-IV0,EXOTIC-TW0,EXOTIC-TW1,EXOTIC-FOUR0}.

As some fraction of o-Ps inevitably results in 'pick-off' annihilations 
due to collisions with atomic 
electrons of the target material, the observed o-Ps decay rate 
$\lambda_{obs}$ is a sum of the intrinsic o-Ps 
decay rate $\lambda_{{\rm o}\mbox{-}{\rm Ps}}$ and the pick-off 
annihilation rate into $2\gamma$'s, 
$\lambda_{pick}$, i.e.,

\begin{equation}
\lambda_{obs}(t)=\lambda_{3\gamma}+\lambda_{pick}(t).
\end{equation} 

Contributions from exotic decays are assumed to be zero because 
they are confirmed to be less than 
$200~ppm$ \cite{EXOTIC-LL1,EXOTIC-LL0,EXOTIC-SL0,EXOTIC-SL1,EXOTIC-SL2,EXOTIC-UB0,EXOTIC-IV0,EXOTIC-TW0,EXOTIC-TW1,EXOTIC-FOUR0}. $\lambda_{pick}(t)$ is 
proportional to the rate of o-Ps collisions with the target materials, i.e.; 
$\lambda_{pick}=n\sigma_a v(t)$, where $n$ is 
the density of the target, $\sigma_a$ is the annihilation cross 
section, and $v(t)$ the time dependent velocity of o-Ps. 
Due to the thermalization process of o-Ps, 
this necessitates expressing 
$\lambda_{pick}$ as a function of time whose properties are dependent on 
the surrounding materials.

In previous measurements \cite{GAS89,CAV90}, $\lambda_{obs}$'s were measured 
by varying the 
densities of the target materials ($n$) or size of the cavities. 
For the uniform distribution of o-Ps in the cavity, $\lambda_{pick}$ 
is depending on the size and shape of the cavity due to the relation,
$v(t)=(\bar{v(t)}/4)S/V$, where $\bar{v(t)}$ is 
the average o-Ps velocity, $S$ is the cavity surface area, 
and $V$ is the volume of the cavity.  

The extrapolation to zero gas density or infinite cavity size 
was expected to yield the decay rate in a vacuum, 
$\lambda_{3\gamma}$, under 
the assumption of quick thermalization 
with constant o-Ps velocity. However, this assumption contains 
a serious systematic error as pointed out in 
reference \cite{THERM_ORIG} in that it would make the obtained 
decay rate larger than the true value; 
i.e., the time-spectrum fitting was carried out using a simple 
exponential function, $dN(t)/dt=N_0\exp(-\lambda_{3\gamma}~t)+C$, 
where $C$ represents the flat spectrum of accidental events. 
In order to cleanly eliminate the contribution from the prompt 
annihilations and fast p-Ps decay component, 
the decay spectrum needs to be fitted for times after 
these components decay away. 
A somewhat fast, beginning time of the fitting range was adopted to obtain 
the high statistics: the cavity experiment typically used 
$t_{start}=170~ns$ for all runs \cite{CAV90}, while the gas 
experiment \cite{GAS89} used $t_{start}=180~ns$ for all gases 
except the lowest pressure $Ne$ and neopentane runs. Because it takes 
several hundred $nsec$ for o-Ps to thermalize via elastic 
collisions with surrounding molecules to an ultimate energy of about 
$0.03~$eV depending on the type of target material 
\cite{MICHIGAN-THERM,MICHIGAN-THERM2}, such fast $t_{start}$ tends 
to result in larger values of $\lambda_{3\gamma}$. 
Moreover, in the cavity experiment, o-Ps is not well thermalized 
due to the small escape and collision rates. 

The new technique called {\it the direct $2\gamma$ correction method}
was introduced to overcome the problem regarding the extrapolation. 
The following section briefly explains its concept and procedure.   
Once a precise thermalization function is obtained, $\lambda_{pick}(t)$ will 
contain all information about the process. The population of o-Ps at time $t$, 
$N(t)$ can be expressed as \begin{equation}
N(t)=N_0' \exp\left(-
\lambda_{3\gamma}\int^{t}_0\left(1+\frac{\lambda_{pick}(t')}{\lambda_{3\gamma}}\right)dt'\right).
\end{equation}

The energy distribution of photons from the 3-body decay is continuous 
below the steep edge at 511~keV, whereas the pick-off annihilation 
is 2-body which produces a 511~keV monochromatic peak. Energy and 
timing information are simultaneously measured with high-energy 
resolution germanium detectors such that 
$\lambda_{pick}(t)/\lambda_{3\gamma}$ can be determined from the 
energy spectrum of the emitted photon. Providing the ratio is 
determined as a function of time, the intrinsic decay rate of o-Ps, 
$\lambda_{3\gamma}$, can be directly obtained by fitting the 
observed time spectrum. 
The direct $2\gamma$ correction method makes any extrapolation procedure 
unnecessary and precluding use of the hypothesis of linear dependence 
at vacuum limit or o-Ps behavior at the material surface. 
\section{Experiment}\label{sec:experiment}
The direct 2$\gamma$ correction method first applied in 1995 
\cite{ASAI95} provided consistent results with QED predictions, 
yet several problems remained: (i) accuracy was $400~ppm$, being 
worse than those of the other experiments \cite{GAS89,CAV90}, 
(ii) decay rates systematically increased before $t_{start}=200~ns$ 
due to an unknown reason, and therefore to remove this uncertainty, 
final results were obtained using data after $220~ns$, and (ii) 
systematic error regarding the Stark effect was not 
estimated. Improving the method by considering these problems will be 
described in the later sections. 
\begin{figure}[t!]
\vspace{0.2pc}
  \begin{center}
    \includegraphics[width=14pc, angle=-90,keepaspectratio]{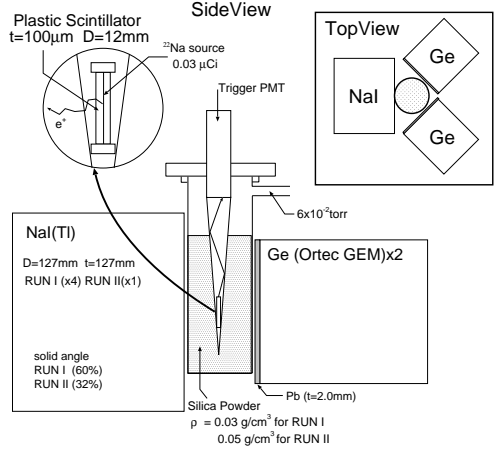}
  \end{center}
\caption{Schematic diagram of experimental setup. $D$, $t$ represent 
the diameter and the thickness (longitudinal length) of each 
cylindrical component respectively. The density, $\rho$, and the 
other characteristics of the silica powder are summarized in 
Table~\ref{table:powder}. 
Upper right figure illustrates the geometrical location of each 
component viewed from top.}
\label{fig:setup}
\end{figure}
We employed a ${}^{22}{\rm Na}$ positron source (dia., $2~mm$) with 
approximate 
strength of $0.03\mu Ci$, being sandwiched between two sheets of plastic 
scintillators (NE104) and held by a cone made of aluminized mylar 
(Fig.~\ref{fig:setup}). The scintillators and mylar were $100~\mu m$- and 
$25~\mu m$-thick, respectively, corresponding to a total area density of about 
$17~mg/cm^2$. The cone is situated at the center of a cylindrical 
$50~mm$-diameter vacuum container made of $1~mm$-thick glass, being 
filled with ${\rm 
SiO_2}$ powder and evacuated down to $5\times10^{-2}$ Torr. Two different 
types of ${\rm SiO_2}$ powder (Nippon Aerosil Ltd.) were used 
(Table~\ref{table:powder}), with the biggest difference being in 
the mean distance between grains ($340$ vs. $200~nm$) such that 
different pick-off ratio would be obtained. Using these 
powders, two 6-month runs were performed (RUN1 and 2). 
To remove water molecules absorbed on the grain surface, the 
powder was heated just before evacuation.

\renewcommand{\arraystretch}{1.00}
\begin{table}[htb]
\begin{center}
\begin{tabular}{crr}\hline
                   & RUN1 & RUN2 \\ \hline\hline
primary grain size (nm)&        7  &       7 \\ 
surface area  ($m^2/g$)&   $300\pm30$ & $260\pm30$ \\
density ($g/cm^3)$       &   0.03 & 0.05 \\ 
mean distance between grains (nm)     &   340  & 200 \\ 
surface & hydrophile & hydrophobe \\ \hline
\end{tabular}
\caption{Characteristics of ${\rm SiO_2}$ powders used in the measurements.}
\label{table:powder}
\end{center}
\end{table}
While traveling through the scintillators, the source-emitted 
positrons deposit energy from $50-100$ keV which produces 
scintillation photons transmitted to the trigger PMT 
(Hamamatsu H3165-04). The magnetic field around the assembly 
is measured to be $0.5\pm0.1~Gauss$, equivalent to terrestrial 
magnetism which barely contributes to the mixing between o-Ps and 
p-Ps estimated to be $3\times10^{-11}$. The observed o-Ps is therefore 
regarded as a pure o-Ps sample. 

Two high-purity coaxial germanium detectors (Ortec GEM38195), 
referred to as Ge0 and Ge1, precisely measured the thermalization 
process, respectively having crystal sizes of 58.3 and 60.3 mm in 
dia. and 73.8 and 67.4 mm in length, and a solid angle of 5.7 and 
6.0\% of $4\pi$. Energy resolutions were measured using several 
line $\gamma$ sources, with typical resultant values of $0.53$ 
and $0.64~{\rm keV}$ in sigma for Ge0 and Ge1, respectively, 
at 514~keV. Lead sheets 2.0-$mm$-thick were placed in front of each 
detector for suppressing contributions from simultaneous low-energy 
$\gamma$ hits from the 3$\gamma$ decay of o-Ps.

Four large cylindrical NaI(Tl) scintillators (Scionix 127A127/5; 
127~$mm~(\phi)$ $\times$ 127~$mm~(t)$) 
simultaneously measured the time and energy information from 
each decay. Due to their higher efficiency 
and faster time response, the time spectrum can be fit to 
the o-Ps decay rate determination. At 662~keV, 
their energy resolution is typically 22.0, 24.4, 21.4, and 32.0~keV.

A new time-to-digital converter (TDC) was employed, jointly developed 
with the High Energy Accelerator Research Organization (KEK) and customized 
for the present measurement. This direct clock-counting type TDC, referred 
to as the 2-GHz TDC, has a time-resolution of $0.5~nsec$ with
known accuracy of 1-$ppm$, 
equipped with 8 channels being able to accommodate multiple signals. 
The time range for each channel is $32~\mu s$,  
and the integral non-linearity (INL) is expected to
be extremely small at $<15~ppm$. To provide a systematic check, we 
used a 200-MHz internal clock-based TDC (Hoshin C006) with 
5~$ns$ time resolution.
\begin{figure}[t!]
\vspace{0.2pc}
  \begin{center}
    \includegraphics[width=12pc, angle=-90,keepaspectratio]{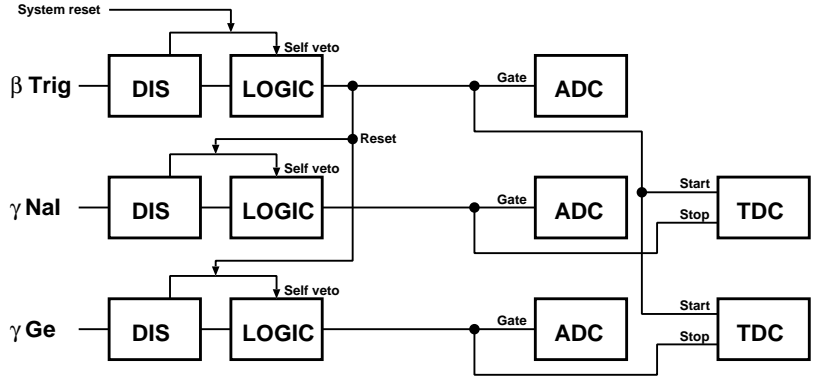}
  \end{center}
\caption{Simplified scheme of the trigger system. In figure, DIS and LOGIC
represent the discriminator and the logic unit components respectively. 
Descriptions of signal flows are found in the text.}
\label{fig:trigger}
\end{figure}

The electronic system consists of three sections: the trigger, 
NaI(Tl), and Ge sections. Each has a latch veto 
structure to prevent overlapping events. The main latch of the 
trigger section is set by the signals 
themselves and released with system reset. To ensure precise 
measurement, the latches of the NaI(Tl) and 
Ge sections are also set by themselves and released by reset 
signals from the trigger section. The trigger 
PMT signal is fed into a fast leading discriminator whose 
output provides common start signals for the 
TDCs as well as their reset signals. Each output signal 
from the photo-multipliers to the NaI(Tl)s is fed into 
three ADCs and a fast leading edge discriminator that provides 
stop signals for the TDCs and gate timing for the ADCs. 
One ADC with a 3-$\mu s$ gate width, called the NAI-WD 
(Lecroy 2249W), measures the whole charge for the duration 
of the signal, with slewing correction being carried out using energy 
information. The other two ADCs eliminate pile-up events 
at the tail of the signal and base-line fluctuations, 
respectively. The NAI-NW (mainly Lecroy 2249W) ADC measures 
the signal charge with a narrower gate 
width (250\ $ns$ for RUN1 and 400\ $ns$ for RUN2) comparable 
to the intrinsic decay time constant of 
the NaI(Tl) scintillator, while the NAI-BS (Lecroy 2249A) 
ADC measures the base-line condition of the 
signal (180\ $ns$ width) just prior to the event. All gates 
for these ADCs are individual and gate timings 
are optimized for signal timings.

Signal outputs from the Ge detectors are used for 
timing measurements and obtaining precise energy 
information. One signal is fed into a fast-filter 
amplifier (FFA)(Ortec 579) whose output is used as the stop 
signal for the TDCs, two auxiliary ADC signals, and three 
discriminators whose thresholds are set at $-50, -100$, and 
$-150$~mV. Each FFA output is fed into another high-resolution 
TDC having fine resolution of $250~ps$. Signal timings are 
determined utilizing signal shapes calculated from different threshold 
discriminating times and extrapolating to the intrinsic timing 
\cite{D-thesis}. Good time resolution of 
$4~ns$ is obtained, as is efficient rejection power 
for slow rise signal components known to disturb Ge 
timing spectroscopy. Similar to the NaI(Tl) detectors, 
the two auxiliary ADCs effectively reject pile-up 
events by measuring the signal charge with a narrower 
gate (2.0(1.3)$~\mu s$ width for RUN2(1)) and 
earlier timing gate (200~ns width just prior to 
the signal). Precise time and energy resolutions can be 
obtained with this rejection scheme. 
The other output from each Ge detector is independently 
fed into a SHAPER spectroscopy amplifier (Ortec 
673). Amplified with a 6-$\mu s$ time constant, 
the uni-polar output is provided to a Wilkinson type peak-holding 
ADC called GE-WD that provides a precise energy spectrum. 
For RUN1 and RUN2, $2.7\times10^9\ \beta^+$ and 
$2.0\times10^9\ \beta^+$ events were recorded, respectively. Each run had a 
total data acquisition period of about six months. 
Area temperature was maintained within $\pm0.5{}^\circ C$ 
to ensure stability of the amplifiers, ADCs, and TDCs.
\section{Analysis and discussion}\label{sec:analysis}

\begin{figure}[t!]%
\vspace{0.2pc}
  \begin{center}
    \includegraphics[width=15pc,keepaspectratio]{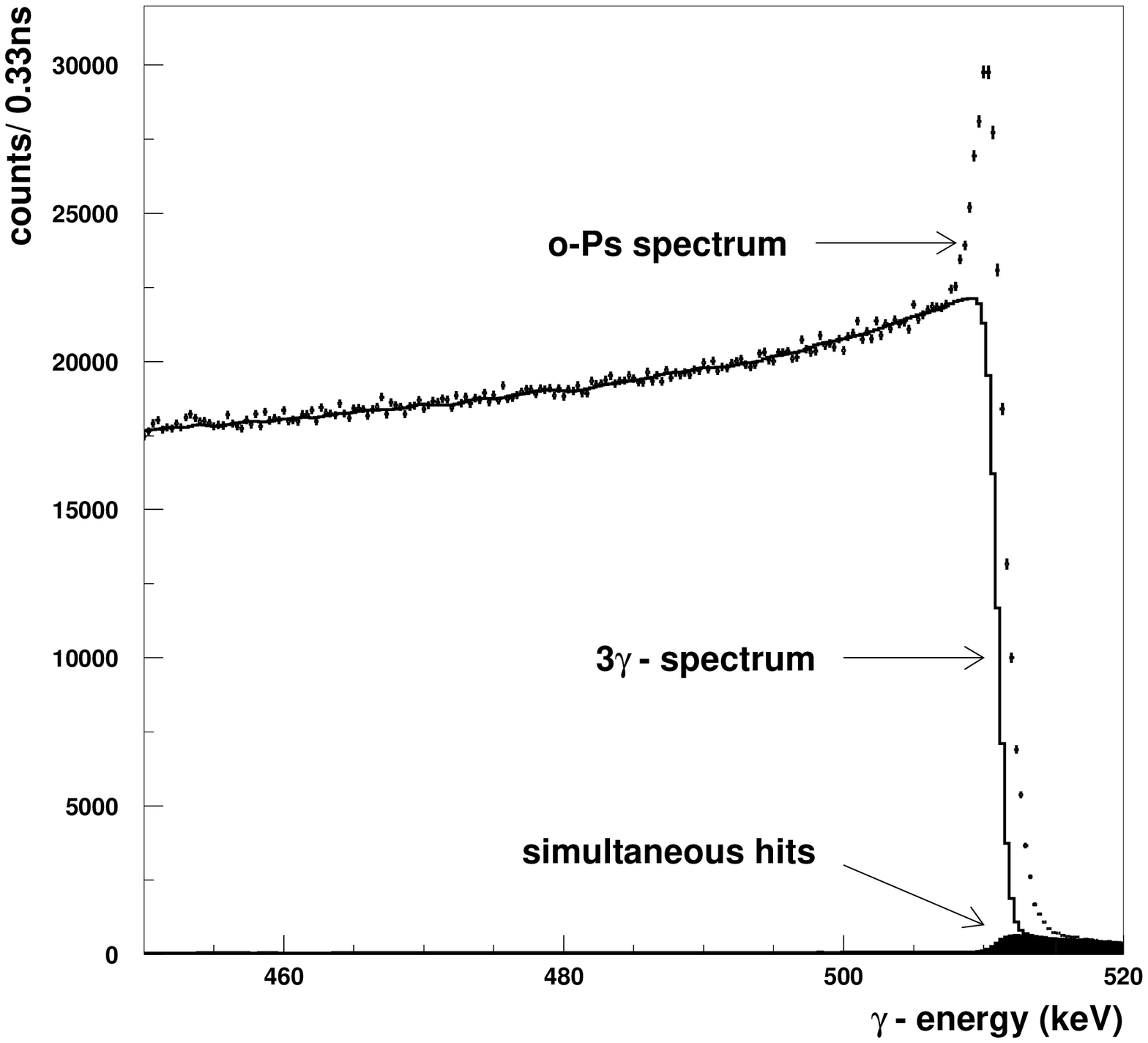}
    \includegraphics[width=15pc,keepaspectratio]{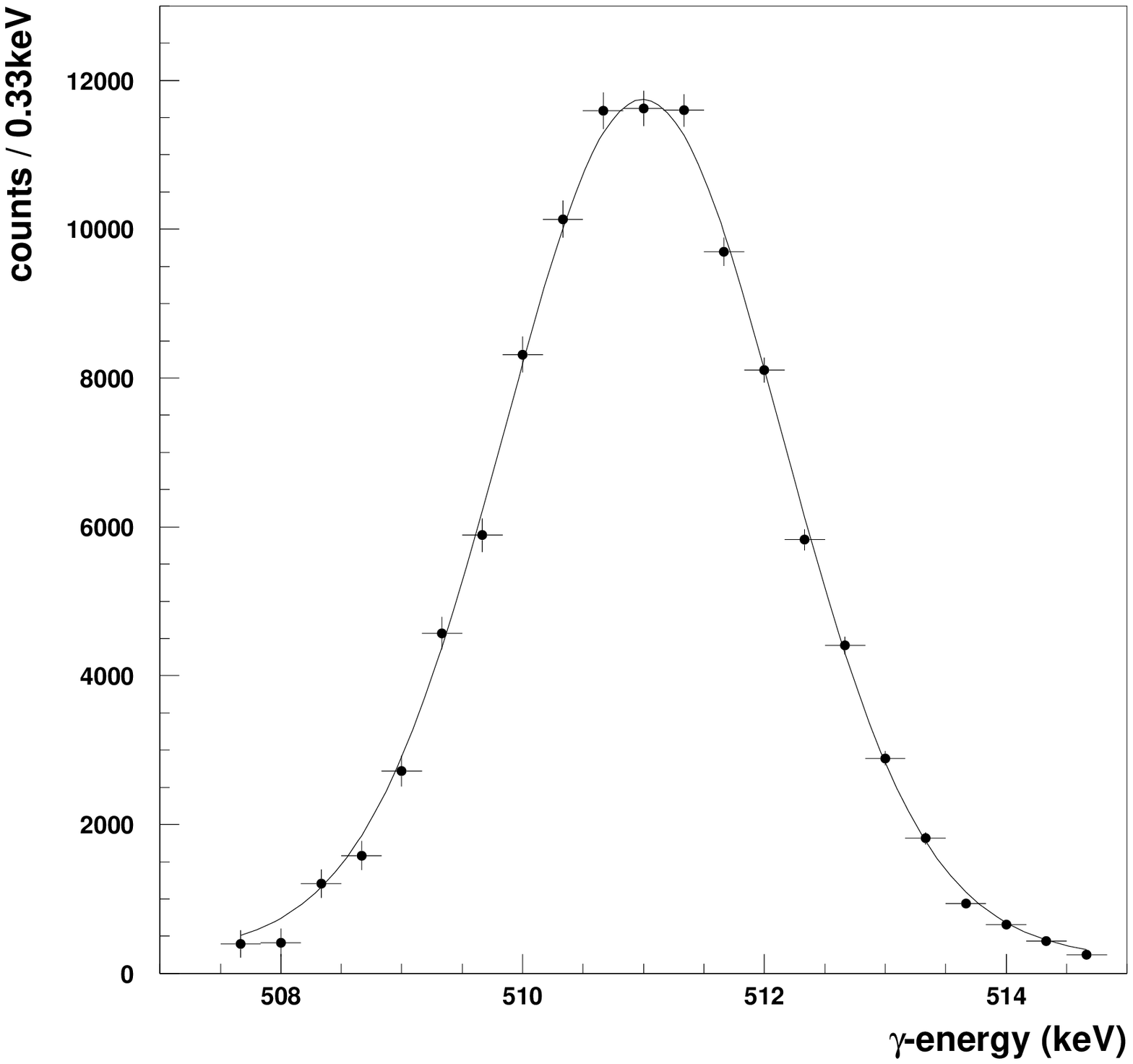}
  \end{center}
\caption{(a) Energy spectrum of o-Ps decay $\gamma$'s obtained by Ge 
detectors for RUN1. Dots with error bars represent data points in a 
time window of $150-700~ns$. The solid line shows the $3\gamma$-decay 
spectrum calculated by Monte Carlo simulation. Shaded area indicate 
simultaneous hits 
of two low-energy $\gamma$'s from a 3-$\gamma$ decay estimated 
using the same simulation. 
(b) Pick-off spectrum obtained after subtracting the $3\gamma$ contribution 
from the o-Ps spectrum for RUN1. The solid line represents the fit result.}
\label{fig:ops-3gamma}\
\end{figure}

The ratio $\lambda_{pick}(t)/\lambda_{3\gamma}$ is 
determined using the energy spectrum measured by 
the Ge detectors. The energy spectrum of the o-Ps 
decay sample, referred to as the {\it o-Ps spectrum}, is 
obtained by subtracting accidental contributions from 
the measured spectrum. The $3\gamma$-decay 
continuum spectrum is calculated using Monte Carlo 
simulation in which the geometry and various 
material distributions are reproduced in detail. 
Three photons are generated according to the 
$O(\alpha)$ corrected energy spectrum \cite{Adkins_Private}, 
followed by successive photo-electric, 
incoherent, and coherent interactions with 
materials until all photon energy is either deposited or escapes 
from the detectors. The response function of the detectors 
is determined based on the measured spectrum of 
monochromatic $\gamma$-rays emitted from 
${}^{152}{\rm Eu}$, ${}^{85}{\rm Sr}$, and 
${}^{137}{\rm Cs}$, with this function being 
used in the simulation. We refer to the obtained spectrum as 
the {\it $3\gamma$-spectrum} which is normalized to the 
o-Ps spectrum with the ratio of event numbers 
within the energy domain 480-505~keV. 
The region is chosen to minimize, the Compton scattering effect from 
pick-off events, dependence of the detector efficiencies on energy, 
and the $O(\alpha)$ correction to the $3\gamma$-spectrum.
Figure \ref{fig:ops-3gamma}(a) shows good agreement 
between the o-Ps spectrum and $3\gamma$ spectrum below 
508~keV, where the pick-off annihilation peak is evident 
at the edge of the $3\gamma$-decay continuum. 
It should be noted that the lead sheets in front of 
the Ge detectors effectively 
suppress the contribution from events in which two 
low-energy $\gamma$'s emitted from a $3\gamma$-decay 
simultaneously hit one detector and that the $3\gamma$-spectrum 
well reproduces such simultaneous events.

Figure \ref{fig:ops-3gamma}(b) shows an enlarged view of 
the observed o-Ps spectrum after subtracting 
the $3\gamma$-spectrum, where good agreement with a 
detector response function \cite{PICK-0,PICK-1} 
is present. The centroids of the peak spectrum are 
obtained as $510.997~^{+0.003}_{-0.024}~\mbox{(keV)}$ 
for RUN1 and $510.995~^{+0.008}_{-0.059}~\mbox{(keV)}$ for RUN2. 
Good consistency with $511.0$~(keV) indicates successful 
subtraction of the $3\gamma$ contribution such that 
the resultant peak can be regarded as pure pick-off 
annihilation samples. Obtained ratios of 
$\lambda_{pick}/\lambda_{3\gamma}$ at a typical time 
window of $150-700~ns$ are $0.01049(8)$ and 
$0.01582(9)$ for RUN1 and RUN2, respectively. The weighted 
mean time within the range for each run is 
$287.5~ns$ (RUN1) and $284.3~ns$ (RUN2). The large 
difference between the two runs originates from a 
different mean distance of powder grains 
(Table~\ref{table:powder}) since the lambda ratio is a function of time.

\begin{figure}[t!]%
\vspace{0.2pc}
  \begin{center}
    \includegraphics[width=20pc,keepaspectratio]{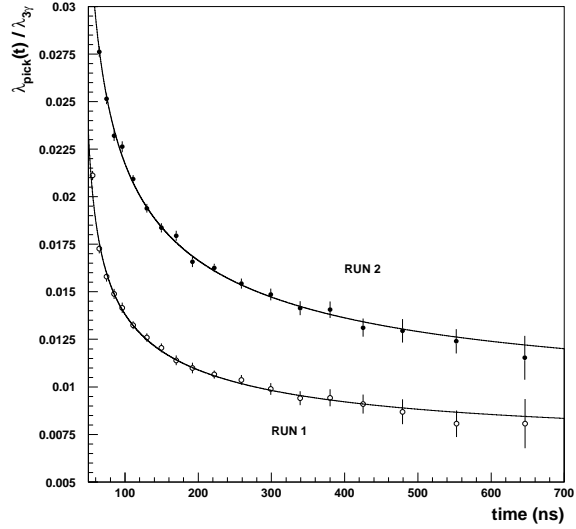}
  \end{center}
\caption{The ratio $\lambda_{pick}(t)/\lambda_{3\gamma}$ 
are plotted as a function of time. Open circles are data 
points for RUN1 and closed circles for RUN2. Solid lines 
represent best fit results obtained using Eq.~\ref{eq:therm2}. 
}
\label{fig:pick_ratio}
\end{figure}

Figure~\ref{fig:pick_ratio} shows results when the procedures 
described above for a typical time window 
were applied to multiple, narrower time windows. 
Since the fractional energy loss of o-Ps per collision with 
${\rm SiO_2}$ powder and the collision rate are both dependent 
on its energy, the time dependence of the 
average kinetic energy of o-Ps at time $t$, 
$\overline{E(t)}$ can be derived from the Boltzmann equation 

\begin{equation}
\frac{d}{dt}\overline{E(t)}=-\sqrt{2m_{Ps}\overline{E(t)}}\left(\overline{E(t)}-\frac{3}{2}k_BT\right)\sum^\infty_{j=0}a_j\left(\frac{\overline{E(t)}}{k_BT}\right)^{j/2},
\label{eq:therm1}
\end{equation}

where $m_{Ps}$ is the mass of o-Ps, $T$ room temperature, 
and $k_B$ the Boltzmann constant. The last 
term, the momentum transfer cross section of 
${\rm SiO_2}$ is expanded in terms of $\overline{E(t)}$, i.e., 
the coefficients $a_j$ represent the effect of effective mass 
at the surface of the ${\rm SiO_2}$ grain and 
mean distance between the grains. Since the 
pick-off rate is proportional to the average velocity of the o-Ps, 
the ratio $\theta(t)\equiv\lambda_{pick}(t)/\lambda_{3\gamma}$ 
can be expressed by a differential equation \cite{SIO2-1,SIO2-2}, i.e.,

\begin{equation}
\frac{d}{dt}\theta(t)=-C\left(\theta(t)^2-\theta_\infty^2\right)\theta(t)^{2\beta},
\label{eq:therm2}
\end{equation}

where $C$ is a constant, $\theta_\infty\equiv\theta(t\rightarrow\infty)$, 
and the last summation term in Eq.~\ref{eq:therm1} is 
replaced with an arbitrary real number $\beta$.
The measured $\lambda_{pick}(t)/\lambda_{3\gamma}$'s are fit with 
Eq.~\ref{eq:therm2}, and the first-order differential equation 
is numerically solved using the Runge-Kutta method. 
The data point at a time window of $40-45~ns$ is used for a fixed 
initial value. Fitting is carried out using four parameters: $C$, 
$\theta_{\infty}$, $\beta$, and the deviation from fixed 
point $\delta$ found to be $-0.27\times10^{-4}$ and 
$-0.32\times10^{-4}$ for RUN1 and RUN2, respectively. 
These small values indicate that the fitted values rarely 
depend on the initial condition of the Runge-Kutta method. 

Figure~\ref{fig:pick_ratio} shows best fit results 
using the MINUIT library \cite{MINUIT}, where the 
pick-off rate cannot be assumed as constant, even in ${\rm SiO_2}$ 
powder where the collision rate is expected to be higher. 
Table~\ref{table:pick_fit_result} gives resultant values 
for these fitting parameters.
\begin{table}[htb]
\begin{center}
\renewcommand{\arraystretch}{1.2}
\begin{tabular}{ccc}\hline\hline
Parameters & RUN 1 & RUN 2 \\ \hline
$C$ 
& $\left(0.166~^{+0.043}_{-0.045}\right)\times10^{-2}$ &
$\left(0.522~^{+0.016}_{-0.014}\right)\times10^{-2}$\\
$\theta_\infty$ & 
$\left(0.761~^{+0.090}_{-0.17}\right)\times10^{-2}$ &
$\left(0.999~^{+0.12}_{-0.22}\right)\times10^{-2}$\\
$2\beta$ & 
$2.64{~}^{+0.35}_{-0.30}$ & 
$2.12{~}^{+0.25}_{-0.22}$ \\
$\chi^2$/d.o.f & 9.53/(23-4) & 12.27/(23-4) \\ \hline\hline
\end{tabular}
\caption{Summary of $\lambda_{pick}(t)/\lambda_{3\gamma}$ fitting results.}
\label{table:pick_fit_result}
\end{center}
\renewcommand{\arraystretch}{1.0}
\end{table}

\begin{figure}[t!]%
\vspace{0.2pc}
  \begin{center}
    \includegraphics[width=20pc,keepaspectratio]{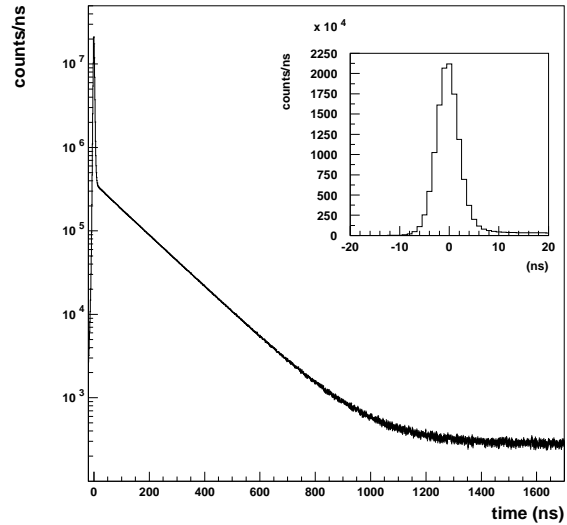}
  \end{center}
\caption{Time spectrum of NaI(Tl) scintillators for RUN 1 
within an energy window of $370-440$~keV. The inset shows an 
enlarged view of the prompt peak, where good time 
resolution of $\sigma=2.2~ns$ is obtained.}
\label{fig:nai_timespectrum}
\end{figure}
Figure~\ref{fig:nai_timespectrum} shows the time spectrum of 
NaI(Tl) scintillators for RUN 1 with an energy window 
of $370-440$~keV, where a sharp peak in prompt positron 
annihilation is followed by the exponential decay curve 
of o-Ps and then the constant accidental spectrum. 
The o-Ps curve is widely observed over about $1.2~\mu s$ 
corresponding to about eight times the o-Ps lifetime, 
being partly due to the use of a weak positron source 
($0.03~\mu Ci$) and good suppression of accidental contributions 
by selecting the $\gamma$-energy. To effectively eliminate 
pile-up events, a base-line cut condition was applied, 
and to further reject pile-up events, events with small 
differences between two ADC values (wide and narrow gates) were selected.

After selecting events in the energy window of $370-440$~keV, 
we fit resultant time spectrum using the least square method, i.e.,
\begin{equation}
N_{obs}(t)=\exp(-R_{stop}t)\left[\left(1+\frac{\epsilon_{pick}}
{\epsilon_{3\gamma}}\frac{\lambda_{pick}(t)}{\lambda_{3\gamma}}\right)N(t)
+C\right], 
\label{eq:tspec_obs}
\end{equation} 
where $\epsilon_{pick}$ and $\epsilon_{3\gamma}$ are 
the detection efficiencies for pick-off annihilations and 
$3\gamma$ decays. $R_{stop}$ in the function is the measured stop 
rate of the NaI(Tl) scintillators, representing the fact that time 
interval measurement accepts only the first $\gamma$ as a stop signal. 
$\lambda_{pick}/\lambda_{3\gamma}$ is about $1\%$ due to 
the low-density of the ${\rm SiO_2}$ powder, i.e., the 
ratio of error propagation to decay rate is suppressed 
by a factor of 100. The value of $\epsilon_{pick}/\epsilon_{3\gamma}$ 
is estimated using Monte Carlo simulation, yielding 
$0.0651(7)$ and 0.0607(6) for RUN 1 and RUN2, respectively. 
The time spectrum is corrected in terms of the $R_{stop}$ (order of 
1~kHz) before fitting.
Fitting stop time is fixed at $5000~ns$. 

\begin{figure}[t!]%
\vspace{0.2pc}
  \begin{center}
    \includegraphics[width=20pc,keepaspectratio]{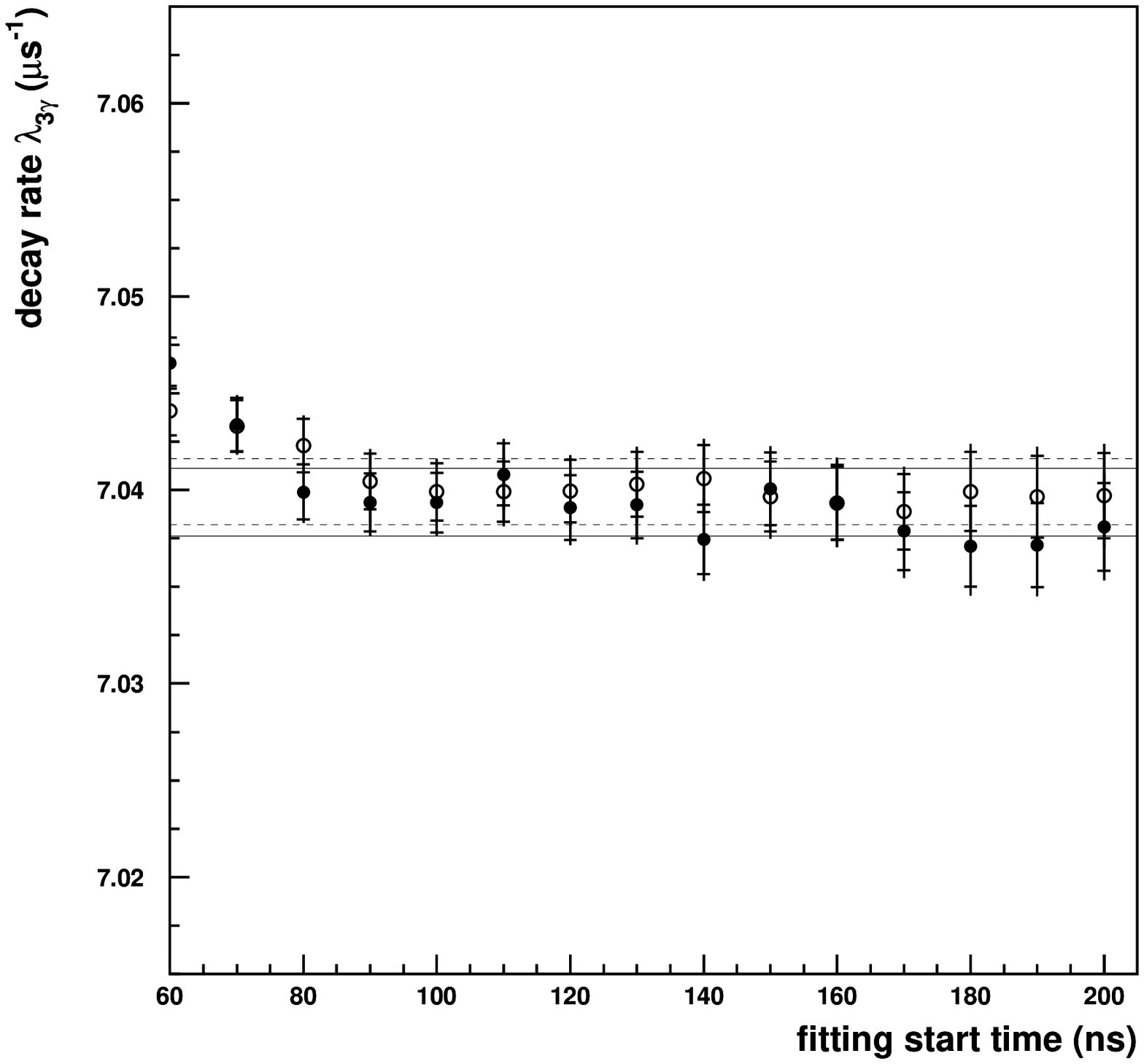}
  \end{center}
 \caption{Decay rates as a function of fitting start time. 
Small horizontal lines on the error bars represent the size 
of statistical errors solely due to fitting, while 
vertically extending bars include propagated errors from 
$\lambda_{pick}(t)/\lambda_{3\gamma}$ determination. 
Open and closed data points indicate values for RUN1 and RUN2, 
respectively. Dashed and solid lines show one standard deviation 
obtained at $t=100~ns$ for RUN1 and RUN2, respectively.}
\label{fig:fit_result}
\end{figure}

Figure~\ref{fig:fit_result} shows obtained fitting as a 
function of fitting start time for both runs, where 
values are stable with respect to fitting start time 
except those before $80~ns$. Since the fitting $\chi^2$'s 
for both runs rapidly increase before $90~ns$ due to the 
tail effect of prompt events, values at 
$t=100~ns$ are taken as the final results. 
The reduced $\chi^2$'s at $t=100~ns$ are $1.006$ and 
$1.051$ for RUN 1 and RUN 2, respectively. Note the 
good correspondence between runs. Since no large 
difference in thermalization properties are apparent 
in Fig.~\ref{fig:fit_result}, this indicates that our method correctly takes 
into account thermalization. The obtained decay rates are 
$\lambda_{3\gamma}=7.03991\pm0.0017(stat.)~\mu s^{-1}$ for RUN 1 and 
$\lambda_{3\gamma}=7.03935\pm0.0017(stat.)~\mu s^{-1}$ for RUN 2. 

\section{Discussion of systematic errors}\label{sec:systematics}
\renewcommand{\arraystretch}{1.00}
\begin{table}[htb]
\begin{center}
\begin{tabular}{lrr}\hline\hline
Source of Contributions&RUN1 ($ppm$)& RUN2 ($ppm$) 
\\\hline\hline
TDC module dependence & & \\
\quad-- Calibration & $<1$ & $<1$ \\
\quad-- Stability   & $2\sim3$ & $2\sim3$  \\
\quad-- Integral Non Linearity & $<~15$ & $<~15$ \\
\quad-- Differential Non Linearity & Negligible & Negligible \\\hline
Cut condition dependence & & \\  
\quad-- Base Line Selection &$-17$ and $+89$ & $-11$ and $+23$
 \\
\quad-- WD-NW condition & $-6$ and $+45$ & $-20$ \\ \hline
Monte Carlo dependence & & \\
\quad-- Normalization & $\pm99$ & $\pm113$ \\
\quad-- Relative efficiency  &  &  \\ 
\hspace*{15mm}of NaI(Tl) scintillator & $+7$ & $+7$\\
\quad-- Inhomogeneity of ${\rm SiO_2}$ powder & $<\pm55$ & $<\pm88$ \\\hline
Other Sources & & \\
\quad-- Zeeman effect & $-5$ & $-5$ \\
\quad-- Three-photon annihilation & $-26$ & $-42$ \\
\quad-- Stark effect & $-21$ & $-6$ \\
\quad-- $n=2$ excited state 	& $+19$ & $+19$\\\hline
Total  & $-120$ and $+153$ &  $-151$ and $+148$ \\ \hline\hline
\end{tabular}
 \caption{Summary of systematic errors. See text for details of each item.}
\label{table:systematic}
\end{center}
\end{table}
The sources of systematic errors are summarized in 
Table~\ref{table:systematic} and each item is depicted in the following 
discussions. 

TDC module related errors:
Calibration accuracy ($< 1~ppm$) and stability of 
the calibration oscillator ($2\sim3~ppm$) are known as 
they are product specifications, while integral ($< 15~ppm$) 
and differential (negligible) non linearities are 
evaluated using basic performance of the 2-GHz TDC.
Cut condition errors:
Systematic errors for base-line selections and pile-up 
rejections were assessed by changing the cut criteria. The 
base-line cut was intentionally changed from $\pm3\sigma$ to 
$\pm5\sigma$ where the $\sigma$ represents standard deviation of 
the base-line distribution. The decay rate results are 
compared with the nominal result with $\pm4\sigma$, and systematic 
error is estimated as $-17~ppm, +89~ppm$ for RUN1 and $-11~ppm, 
+23~ppm$ for RUN2. Regarding the pile-up events rejection, 
the WD-NW cut condition was changed 
from $\pm25 {\rm keV}$ to $\pm40 {\rm keV}$ to 
estimate deviation from the nominal condition which is 
$\pm 30$~keV, and is estimated as $-6~ppm, +45~ppm$ for 
RUN1 and $-20~ppm$ for RUN2. The 
dependence on the energy cut condition 
and trigger PMT threshold are considered to have a negligible effect.

Monte Carlo simulation related errors:
The predominant contribution to total systematic error 
is produced by uncertain normalization. That is, the 
number of pick-off events are determined by subtracting 
the normalized 3-$\gamma$ spectrum of Monte 
Carlo simulation from the o-Ps spectrum, where 
changing the normalization factor affects the 
$\lambda_{pick}(t)/\lambda_{3\gamma}$ values 
and eventually propagates to the final result. Since the 
sharp fall-off of the 3$\gamma$-spectrum at 511~keV 
is solely produced by the good Ge energy resolution 
of $\sigma= 0.5$~keV, this subtraction only affects 
the lower side of the pick-off spectrum such that 
improper subtraction results in asymmetry of the 
pick-off spectrum shape. Comparison of the asymmetries 
of the pick-off peak shape and the prompt peak 
annihilation spectrum is a good parameter for estimating 
this systematic error. The 1~$\sigma$ error is assessed 
as $\pm99~ppm$ for RUN1 and $\pm113~ppm$ for RUN2.

Other MC simulation related errors:
The relative efficiency of the NaI(Tl) scintillators, 
$\varepsilon_{pick}/\varepsilon_{3\gamma}$, has an 
uncertainty of $+5\%$ evaluated using a comparison 
of the energy spectrum of data and that of the 
simulation which reproduces data within an uncertainty 
of 2\% except for the Compton free region where 
the deviation is 5\% ($+7~ppm$). ${\rm SiO_2}$ 
powder density in the MC simulation is conservatively 
changed by $\pm10\%$ although the uniformity is known to be 
within a few \%; a change resulting in an error of $<\pm55$~ppm. 

Stark shift errors:
The Stark shift stretches the lifetime of Ps atoms, 
i.e., a perturbative calculation shows that the shift is 
proportional to a square of the effective electric field 
$E$ such that 
$\triangle\lambda_{3\gamma}/\lambda_{3\gamma}=248\cdot(E/E_0)^2$ 
\cite{STARK}, 
where $E_0 = m_e^2e^5/\hbar^4 \approx 5.14\times10^9~V/cm$. $E$ 
is defined as the root-mean-square electric field sensed by o-Ps 
during its lifetime. Calculations have estimated two contributions 
exist based on measurements of the electrical charge-up on the 
primary grains of silica powders and electrical dipole 
moment on the surface of grains \cite{D-thesis}. 
The charge-up is partly intrinsic depending on powder 
specifications and partly due to positron depositions from 
the $\beta^+$ source. The effect, however, is 
negligible in both cases, i.e., on the level of $10^{-2}~ppm$ 
at most. Silanol functional groups on the 
surface of the powder grain behave as an electrical dipole 
moment creating an effective field around the 
grains. Average densities are known to be as $2.5 /nm^2$ and 
$0.44 /nm^2$ for RUN 1 and 2, respectively. 
Accordingly, the effective field can be analytically calculated 
such that the contribution to the o-Ps decay 
rate is determined to be $-21~ppm$ for RUN1 and 
$-6~ppm$ for RUN2. These estimations were confirmed 
using results from precise hyper-fine-structure (HFS) 
interval measurements of ground state Ps in silica 
powder \cite{HFS}, where the interval is proportional 
to the size of Stark effect. Considering the difference 
in powder densities used, the HFS results are consistent 
with our estimation.

Other sources of systematic errors:
Error contribution due to the Zeeman effect is estimated 
using the measured absolute magnetic field around 
the positronium assembly ($-5~ppm$). Since the 3-$\gamma$ 
pick-off process can only occur at a certain ratio, 
the calculated relative frequency 
$\sigma_{3\gamma}/\sigma_{2\gamma}\sim1/378$ is consistent 
with previous measurements \cite{3-pick}, being $-26~ppm$ for 
RUN1 and $-42~ppm$ for RUN2. The 
probability of the excited state (n=2) of Ps is about 
$3\times10^{-4}$ \cite{EXCITED}, which could make 
the intrinsic decay rate $19~ppm$ smaller due to the 
low decay rate ($\frac{1}{8}\lambda_{3\gamma}$) 

The above discussed systematic errors are regarded as 
independent contributions such that the total 
systematic error can be calculated as their quadratic 
sum, resulting in $-120~ppm$, $+153~ppm$ for RUN1 
and $-151~ppm$, $+148~ppm$ for RUN2.
\section{Conclusions}\label{sec:conclusion}
\begin{figure}[t!]%
\vspace{0.2pc}
  \begin{center}
    \includegraphics[width=17pc,keepaspectratio]{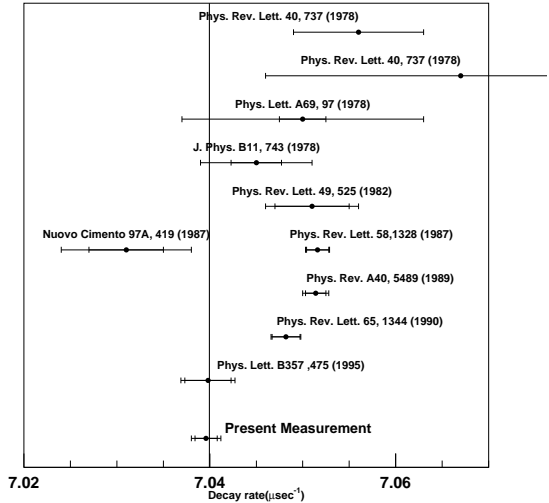}
  \end{center}
 \caption{Historical plot of o-Ps decay rate measurements 
including present results. Vertical line shows the 
$O(\alpha^2)$-corrected NRQED prediction \cite{ADKINS-4}. 
Small vertical lines on the error bars indicate the size 
of errors solely due to statistics 
while error bars represent total ambiguities including systematic errors.}
\label{fig:history}
\end{figure}

The decay rate of o-Ps formed in ${\rm SiO_2}$ was 
measured using a direct $2\gamma$ correction 
method in which the thermalization effect of o-Ps 
is accounted for and integrated into the time spectrum 
fitting procedure. Results were obtained using two runs 
with different types of ${\rm SiO_2}$ powders, i.e., 
$\lambda_{\rm o-Ps}(\mbox{RUN1})=7.03991\pm0.0017(stat.)^{+0.0011}_{-0.0008}(sys.)~\mu s^{-1}$ and 
$\lambda_{\rm o-Ps}(\mbox{RUN2})=7.03935\pm0.0017(stat.)$ ${}^{+0.0010}_{-0.0011}(sys.)~\mu s^{-1}$.
Based on the results of both runs, a weighted average 
gave 
$\lambda_{\rm o-Ps}=7.0396\pm0.0012(stat.)\pm0.0011(sys.)\mu s^{-1}$, 
being 1.8 times more accurate than previous ${\rm SiO_2}$ 
measurement \cite{ASAI95}. 
While this value agrees well with previous result, 
it disagrees with recent high-precision measurements, 
i.e., by $5.6~\sigma$ \cite{GAS89} and 
$3.8~\sigma$ \cite{CAV90}. As illustrated in 
Fig.~\ref{fig:history}, our value agrees well with the 
NRQED prediction corrected up to $O(\alpha^2)$ term 
\cite{ADKINS-4}. In fact, the resultant decay rates 
were confirmed to be independent of fitting start 
time after $t_{start}=90(80)~ns$ for RUN1(2), and 
systematic increase in decay rate observed in previous measurement 
before $t_{start}=200~ns$ \cite{ASAI95} was 
eliminated using the new system; this improvement 
directly contributed to reduce the statistical error. 

Sincere gratitude is extended to Professor Toshio Hyodo, 
Dr. Yasuyuki Nagashima, and Dr. Haruo Saito 
for their valuable suggestions and information 
regarding o-Ps properties and thermalization processes. 
Special appreciation goes to Dr. Masami Chiba for beneficial 
discussions on the o-Ps decay problem and to 
Drs. Masahiro Ikeno and Osamu Sasaki for significant 
measurement improvement using the 2-GHz TDC module.

{\it Appendix}\\
During the submission procedure of this paper, we became aware 
of the recent result using nanoporous silica film by Ann Arbor 
group \cite{michigan2003}.
They obtained the value, 
$\lambda_{\rm o-Ps}=7.0404\pm0.0010(stat.)\pm0.0008(sys.)~\mu s^{-1}$,
which shows complete agreement with our result.


\bibliographystyle{elsart-num}
\end{document}